# Decoupling Urban Food Accessibility Resilience during Disasters through Time-Series Analysis of Human Mobility and Power Outages

Junwei Ma[1*], Bo Li[1], Xiangpeng Li[1], Ali Mostafavi[1]

1 Urban Resilience.AI Lab, Zachry Department of Civil and Environmental Engineering, Texas A&M University, College Station, Texas, United States.

* Corresponding author: Junwei Ma, E-mail: jwma@tamu.edu.

## Abstract

Disaster-induced power outages create cascading disruptions across urban lifelines, yet the timed coupling between grid failure and essential service access remains poorly quantified. Focusing on Hurricane Beryl in Houston (2024), this study integrates approximately 173,000 15-minute EAGLE-I outage records with over 1.25 million revealed visits to 3,187 food facilities across 140 ZCTAs to quantify how infrastructure performance and human access co-evolve. We construct daily indices for outage characteristics (intensity, duration) and food access metrics (redundancy, frequency, proximity), estimate cross-system lags through lagged correlations over zero to seven days, and identify recovery patterns using DTW k-means clustering. Overlaying these clusters yields compound power-access typologies and enables facility-level criticality screening. The analysis reveals a consistent two-day lag: food access reaches its nadir on July 8 at landfall while outage severity peaks around July 10, with negative correlations strongest at a two-day lag and losing significance by day four. We identify four compound typologies from high/low outage crossed with high/low access disruption levels. Road network sparsity, more than income, determines the depth and persistence of access loss. Through this analysis, we enumerate 294 critical food facilities in the study area requiring targeted continuity measures including backup power, microgrids, and feeder prioritization. The novelty lies in measuring interdependency at daily operational resolution while bridging scales from communities to individual facilities, converting dynamic coupling patterns into actionable interventions for phase-sensitive restoration and equity-aware preparedness. The framework is transferable to other lifelines and hazards, offering a generalizable template for diagnosing and mitigating cascading effects on community access during disaster recovery.

# 1. Introduction

The escalating frequency and intensity of climate-induced disasters, such as hurricanes, heatwaves, and wildfires, pose significant threats to urban critical infrastructure systems worldwide, particularly those sustaining essential lifelines like electricity and food supply chains (Feng, Ouyang et al. 2022, Xu, Feng et al. 2024). As cities grow denser and more interdependent, cascading infrastructure failures have become a defining characteristic of disasters, amplifying both physical and social risks (Brunner, Peer et al. 2024). Extreme weather events such as Hurricanes Harvey (2017), Ida (2021), and Beryl (2024) demonstrated how prolonged power outages and supply chain breakdowns can degrade access to basic goods and services, leaving millions without refrigeration, retail access, or reliable distribution (Sciences, Medicine et al. 2020, Feng, Lin et al. 2025, Li, Ma et al. 2025). These disruptions extend beyond immediate physical damage; when basic services falter, urban functionality and community well-being deteriorate, and existing inequalities in access to essential resources are further magnified (Yuan, Farahmand et al. 2023). Ensuring the continuity and equity of accessibility to basic public services (e.g., food) has therefore become a central concern in advancing urban resilience under a changing climate.

Recent research has made significant progress in understanding how disasters reshape human mobility and accessibility to essential services. Leveraging multi-source human mobility and point of interest (POI) data, scholars have captured how residents modify daily visitation behaviors in response to hurricanes, floods, or wildfires. For example, Cai et al. (2024) integrated large-scale mobile phone and precipitation data across 35 Chinese cities to analyze how extreme rainfall reshapes urban mobility, revealing a marked decline in trip frequency and distance as residents concentrated movements within nearby safe zones (Cai, Yang et al. 2024). Ma and Mostafavi (2025) applied high-resolution human mobility data from Hurricane Ida to identify disruptions and recovery in lifestyle activities, showing that visitation frequency to grocery stores and other essential services sharply decreased during landfall and recovered unevenly across communities (Ma and Mostafavi 2025). Li et al. (2025) utilized interpretable machine learning and multi-hazard mobility data from wildfires to predict evacuation behavior, uncovering consistent contraction of travel ranges and prioritization of proximate destinations during high-risk periods (Li, Liu et al. 2025). These studies highlight that disasters consistently trigger localized mobility contractions and adaptive behavioral responses that shape community resilience.

Despite this growing body of work, three critical gaps remain in understanding the interdependent dynamics between infrastructure disruptions and human accessibility. Empirical research on disaster-time food access has traditionally examined either human mobility and consumer behavior or infrastructure performance, but rarely the coupling between the two. Consequently, we have rich descriptions of how people shorten trips, change destinations, or reduce visits under stress, alongside separate accounts of grid failure and restoration, yet little quantified evidence tracing how a lifeline outage such as electricity propagates into measurable losses in food service access. This analytical separation of behavioral and infrastructural domains has left the compound risk pathway that connects power system failure, facility operations, and human access largely uncharacterized in empirical studies of real-world disasters. Furthermore, even when interdependencies are acknowledged, most analyses rely on static or temporally aggregated measures such as event-window totals or weekly averages. These approaches mask short-lived yet policy-relevant dynamics, such as the temporal offsets between the initial shock, infrastructure restoration, and behavioral recovery. Few studies quantify cross-system temporal lags, such as whether power restoration precedes, coincides with, or follows the rebound in food access activity. This temporal blind spot complicates the design of sequenced restoration that anticipates downstream access needs. Moreover, much of the existing research builds on coarse geographic units, underplaying how disparities emerge from the joint interaction of community context (such as road connectivity) and facility-level functionality. Without an approach that links community-level disruption patterns to specific food facilities, it becomes difficult to determine which locations are actually impaired and therefore critical to prioritize in restoring effective access.

This study directly addresses these gaps by empirically coupling high-resolution outage telemetry with fine-grained human visitation to food points of interest during a historic outage event (2024 Hurricane Beryl in Houston). We assemble approximately 173,000 15-minute interval outage records from the U.S. Department of Energy's EAGLE-I™ system and 1.25 million anonymized origin-to-POI trips to 3,187 food facilities across 140 ZIP Code Tabulation Areas (ZCTAs). From this data, we derive daily outage intensity and duration alongside access redundancy, frequency, and proximity metrics, then quantify cross-system timing via lagged correlation and Dynamic Time Warping (DTW)-based K-means algorithm. This design reveals a consistent two-day lag: food access reaches its nadir on landfall (July 8) while outage severity peaks around July 10, with negative correlations strongest at a two-day lag and diminishing by day four. The combination of time-resolved coupling and shape-based clustering yields four compound typologies (power × access) and reveals that road density, more than income, stratifies the severity and persistence of access disruption. Specifically, this study answered three research questions:

**RQ1:** To what extent do cascading infrastructure disruptions (e.g., power outages) affect the temporal dynamics of community food access during disasters?

**RQ2:** How do infrastructural and sociodemographic disparities shape the patterns of coupled outage-food access disruption and recovery across communities?

**RQ3:** Which food facilities can be identified as critical sites for maintaining community food access during and after the disaster?

This work accordingly reframes disaster food security as a multi-system recovery problem where access rebounds not merely when roads clear or demand returns, but when electricity, which enables refrigeration, lighting, and payment systems, is restored in the appropriate places and sequence. Methodologically, the study advances a portable empirical template for interdependency analysis that operates at three critical dimensions: temporal dynamics through daily resolution, spatial scale through ZCTA-to-facility linkages, and decision relevance through identification of 294 critical facilities based on compound exposure, structural connectivity, and operational downtime. The practical implications are substantial. The quantified lag and typologies translate directly into actionable guidance for restoration sequencing, specifically prioritizing feeders that serve zones experiencing both high outage rates and high access disruption. For equity considerations, the findings direct attention to low road-density communities where structural constraints amplify vulnerability. At the facility level, the analysis identifies high-leverage nodes requiring targeted investments in backup power systems and continuity planning. Through measuring the timed coupling of outages and access at operational resolution, this study transforms the field's approach from maintaining parallel narratives about behavior and infrastructure to developing an integrated, actionable science of disaster resilience.

The remainder of this paper is organized as follows. Section 2 reviews relevant literature on disaster-induced food-access disruption and infrastructure interdependencies. Section 3 describes the study context and data sources. Section 4 details the methodology. Section 5 presents the empirical results, and Section 6 discusses the findings, contributions, and limitations of the study.

## 2. Literature review

**2.1. Disrupted food access under disasters**

Food accessibility is a key component of community resilience, reflecting residents' ability to obtain essential food supplies and maintain basic well-being during and after disruptive events. Recent studies have leveraged large-scale mobility, transaction, and retail visitation data to assess how disasters or extreme events affect food access. For example, Wahdat and Lusk (2024) analyze purchasing and spending patterns during Hurricane Ian, illustrating how consumer activity and retail behavior shift markedly in the days surrounding an extreme event (Wahdat and Lusk 2024). Wei and Mukherjee (2025) use mobility-derived measures to capture how actual access to grocery and convenience stores changed across neighborhoods during Winter Storm Uri, highlighting disruptions in people's ability to reach essential food outlets (Wei and Mukherjee 2025). Likewise, Esmalian et al. (2022) track mobility flows during Hurricane Harvey to characterize how disaster conditions altered residents' access to grocery stores and how these changes varied across communities (Esmalian, Coleman et al. 2022).

However, most existing studies conceptualize food access disruption as a direct consequence of physical hazard exposure or behavioral change, without fully considering the operational dependencies of food systems on other critical infrastructures. The implicit assumption is that once transportation and supply routes are passable, food access naturally rebounds. This overlooks how service continuity, such as electricity for refrigeration, lighting, and point-of-sale systems, fundamentally underpins the functioning of food facilities. In practice, disasters rarely affect systems in isolation (Wang, Magoua et al. 2025). Rather, they generate multi-sector disruptions that jointly determine community access to basic needs (Anderson, Brunner et al. 2025). For example, Clark, Perfit, and Reznickova (2024) show that accounting for limited operating hours substantially reduces measured emergency food access, which underscores how facility operations, and the infrastructure services that enable them, are essential components of accessibility during crises (Clark, Perfit et al. 2024). Recognizing these interdependencies requires moving beyond mobility-based or spatial-access models toward frameworks that incorporate the performance of lifeline infrastructures sustaining food systems.

**2.2 Infrastructure interdependencies and cascading disruptions**

Disasters often trigger cascading failures across interconnected infrastructure systems, where the disruption of one lifelines can significantly impair the functionality of dependent sectors including water, transportation, communication, and food supply (Hernandez-Fajardo and Dueñas-Osorio 2013, Rahman, Ingram et al. 2024, Zhang, Mei et al. 2025). In particular, power outages have been identified as a major bottleneck, which impede food access by interrupting the basic functions required to store, handle, and distribute food. Studies have documented that the loss of electricity rapidly compromises refrigeration, leading to widespread food spoilage and heightened food-safety risks during blackout events (Rubin and Rogers 2019). Power loss also disrupts the ability of food retailers to operate, as grocery stores depend on continuous energy supply to preserve perishable inventory and maintain normal business functions, resulting in substantial reductions in available food options when outages occur (Blouin, Herwix et al. 2024). Broader analyses further indicate that extensive or prolonged electricity disruptions can destabilize the food supply chain, affecting food production, processing, and distribution in ways that constrain overall availability (Koetse and Rietveld 2009). Studies on critical infrastructure interdependencies emphasize that such failures amplify social impacts, extending beyond immediate service interruptions to compound vulnerabilities in energy, health, and food security (Henry and Ramirez-Marquez 2016, Zimmerman, Zhu et al. 2018). Despite this recognition, empirical evidence quantifying how power outages translate into measurable food access disruptions remains inadequate. This disconnection limits the ability to capture compound risks that evolve from infrastructure-service coupling during disaster recovery. Addressing this gap requires integrative, data-driven approaches that connect power system performance with real-world measures of food accessibility and mobility.

## 2.3 Dynamic perspectives on compounding disruptions

Beyond recognizing interdependencies among infrastructures, understanding how these relationships evolve over time is also critical. Studies often treats disruptions as static events, summarizing losses or accessibility changes over broad temporal windows and overlooking the day-to-day dynamics of disruption and recovery (Yabe, Rao et al. 2022). Analyses often aggregate observations across extended periods, which can obscure short-term fluctuations that are important for understanding recovery processes (e.g., (Tariverdi, Nunez-Del-Prado et al. 2023)). In reality, infrastructure failures and human behavioral responses unfold continuously: power restoration may not synthesize improvements in mobility, creating temporal misalignments that shape recovery trajectories (He and Cha 2018, Hsu and Mostafavi 2024). Yet few studies have systematically captured these evolving interactions across multiple lifelines, leaving the temporal coupling between infrastructure performance and service accessibility insufficiently understood. A dynamic and fine-grained perspective is therefore essential to reveal how compounding disruptions develop and dissipate over time. Such temporal perspectives highlight that disaster impacts and recovery processes are not uniform but evolve through complex, interdependent pathways. The pace and sequence of restoration can vary widely across systems and locations, reflecting differences in infrastructure capacity, exposure, and community conditions. Understanding these asynchronous and context-dependent patterns remains essential for advancing empirical knowledge of how interconnected infrastructures influence the timing and effectiveness of recovery.

**2.4 Point of departure**

Overall, existing research provides valuable insights into disaster-induced food access disruptions and infrastructure interdependencies, yet several key gaps remain. Most prior studies have examined food access primarily through behavioral or spatial perspectives, without explicitly linking service disruptions to the operational performance of supporting lifeline systems. Empirical work that quantifies how failures in one infrastructure, particularly power, translate into accessibility losses in dependent sectors (e.g., food access) remains limited. Moreover, much of the existing evidence is based on static or temporally aggregated analyses, which overlook the evolving and asynchronous nature of disruption and recovery. Addressing these gaps requires an integrated, dynamic framework that connects infrastructure performance with accessibility outcomes over time. Therefore, this study develops a data-driven approach that combines high-frequency power outage records with fine-resolution human mobility data to quantify the spatiotemporal coupling between energy disruption and food access recovery, thereby advancing understanding of compounding infrastructure-service dynamics during disaster recovery.

# 3. Materials

**3.1 Study context**

This study focuses on the Houston metropolitan area, one of the largest and rapidly growing urban areas in the United States, which experienced severe disruptions following Hurricane Beryl in July 2024. Houston, encompassing Harris County and its surrounding jurisdictions, serves as a critical testbed for examining disaster-induced disruptions due to its high exposure to hurricanes, extreme rainfall, and compound flooding. Past events have repeatedly demonstrated the region's vulnerability to cascading failures across power, transportation, and supply networks, resulting in substantial interruptions to residents' access to essential services such as grocery and food retail facilities (Hong, Bonczak et al. 2021, Lee, Rajput et al. 2022, Li and Mostafavi 2022). Moreover, Houston's large, socio-demographically diverse population and spatially heterogeneous urban form make it a representative environment for investigating the equity and spatial variability of food access during and after natural hazards (Yuan, Farahmand et al. 2023, Wei and Mukherjee 2025).

Hurricane Beryl made landfall on July 8, 2024, near the Texas Gulf Coast with sustained winds exceeding 60-70 mph (National Weather Service 2024). The storm's eyewall directly impacted the Houston metropolitan area, bringing more than 12 inches of rainfall to several locations and triggering extensive urban flooding. These conditions caused widespread transportation interruptions as numerous arterial roads and highways became impassable (Roads&Bridges 7/22/2024). The region's interconnected infrastructure experienced severe cascading failures, with the electric power system hit hardest: more than 2.7 million customers in the Houston area lost power (Michael Zhang 7/11/2024).

The study period spans June 15, 2024, to July 20, 2024, with the baseline period defined as June 15 to July 5, 2024, and the disruption period as July 6 to July 20, 2024. The baseline period represents normal conditions prior to the hurricane's influence. Daily patterns observed during this period were used to establish a baseline for food access behaviors. The disruption period covers the approach, landfall, and aftermath of the hurricane. This study examines 140 ZCTAs within the Houston metropolitan area, enabling a fine-grained assessment of how Hurricane Beryl's compound impacts on infrastructure and mobility jointly influenced spatial and temporal variations in food accessibility.

**3.2 Datasets**

**3.2.1 Power outage data**

High-resolution power outage data were obtained from the U.S. Department of Energy's Environment for Analysis of Geo-Located Energy Information (EAGLE-I™) platform, developed and maintained by Oak Ridge National Laboratory (ORNL). EAGLE-I™ functions as an integrated geographic information system and visualization tool that provides near real-time outage analytics across the United States (Brelsford, Tennille et al. 2024). In this study, we extracted 15-minute interval outage records covering the period from June 15 to July 20, 2024 at the ZCTA level for the Houston metropolitan area. The dataset comprises approximately 173,000 records and each record represents the number of customers without power and timestamp within a given ZCTA. The EAGLE-I™ dataset has been widely applied and validated in prior studies of large-scale power system resilience, providing a robust foundation for quantifying the compound effects of infrastructure failure during extreme events (Li, Ma et al. 2024, Li, Ma et al. 2025, Li, Ma et al. 2025, Ma, Li et al. 2025).

**3.2.2 Human mobility data**

Daily human movements data were derived from Spectus, a large-scale, privacy-preserving GPS location dataset that aggregates anonymized data from over 220 mobile applications across the United States. Spectus data capture movement trajectories from tens of millions of opted-in devices, representing roughly 20 % of the U.S. population (Spectus 2025). The dataset includes anonymized device identifiers, geographic coordinates, and timestamps collected through a compliant, user-consented framework. In this study, each device was assigned a home ZCTA based on its most frequent nighttime location during a multiweek observation window. Daily trips were then aggregated from these home ZCTAs to food-related POIs, generating an origin-destination matrix of food access flows across the Houston metropolitan area. Over the study period, approximately 1.25 million movements to food locations were recorded, forming a dynamic network of food access behaviors (conceptualized in Figure 2b).

Data on food-related POIs were obtained from SafeGraph, which provides comprehensive business information including geographic coordinates, addresses, and North American Industry Classification System (NAICS) codes (SafeGraph 2025). Food-related establishments were identified by filtering NAICS categories corresponding to grocery stores, supermarkets, convenience stores with food sales, specialty food shops (e.g., bakeries, meat markets), quick-service and full-service restaurants, and wholesale food outlets accessible to the public (NAICS Economic Classification Policy Committee). After filtering and spatial verification, 3,187 unique food POIs were identified within the study area (Figure 2a).

### 3.2.3 Auxiliary data

To characterize disparities in food access disruptions, two key metrics were incorporated at the ZCTA level: road density and median household income. These metrics capture dimensions of urban form and social vulnerability, both of which have been shown to influence urban resilience in prior studies (Ma, Blessing et al. 2024, Ma and Mostafavi 2024).

Road density, derived from OpenStreetMap (OSM), quantifies the total length of roadway per square kilometer of land area within each ZCTA (OpenStreetMap 2025). This metric serves as a proxy for local accessibility and the structural characteristics of the built environment. Areas with denser road networks typically exhibit greater connectivity and redundancy, which can facilitate access to essential services during disruptions; conversely, sparsely connected areas may experience constrained mobility and delayed recovery when infrastructure systems fail.

Median household income, obtained from the U.S. Census Bureau's American Community Survey (ACS) five-year estimates (2018-2022), represents the socio-economic status of residents within each ZCTA (U.S. Census Bureau 2025). Income is widely recognized as a determinant of household adaptive capacity, affecting the ability to prepare for, cope with, and recover from infrastructure disruptions (Jing, Heft-Neal et al. 2025). In the context of this study, it is used to differentiate communities' resilience potential and to examine whether lower-income areas experienced disproportionately severe or prolonged food access disruptions following Hurricane Beryl.

## 4. Methodology

Figure 1 illustrated the workflow of this study. First, food access and power outage datasets were collected and processed at the ZCTA level for the Houston metropolitan area. Food access, power outage, and auxiliary metrics, including access redundancy, access frequency, access proximity, outage duration, outage intensity, road density, and median household income, were then developed. Next, time-series clustering algorithm was applied to classify areas into distinct compound impact groups. Subsequent analyses quantified both the temporal lag and spatial disparities between power system disruption and food access recovery. Finally, critical food facilities were identified based on their roles in serving highly impacted or vulnerable communities.

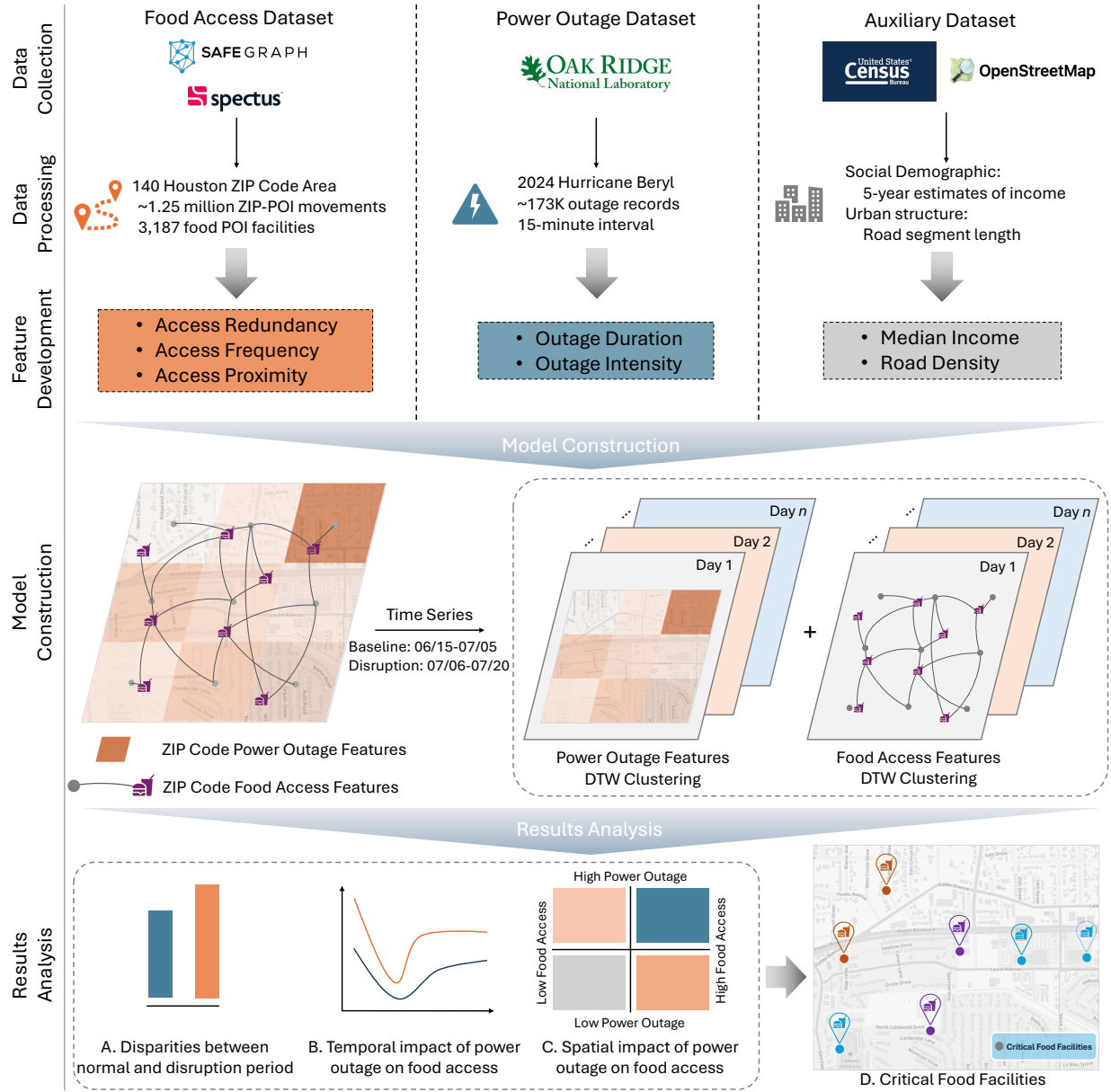

**Figure 1. Study framework linking power outages to food access disruptions during Hurricane Beryl (Houston, 2024).**

## 4.1 Metrics

### 4.1.1 Power outage metrics

From the 15-minute interval power outage dataset, two daily metrics were derived to characterize the magnitude and persistence of power service disruption. Following the approach established in our previous work (Li, Ma et al. 2024, Ma, Li et al. 2025), we excluded intervals in which fewer than 0.1 % of customers were reported without power to minimize noise from minor reporting fluctuations.

**(1) Outage intensity**

The outage intensity represents the average proportion of customers experiencing power loss within a 24-hour period. This metric captures the severity of electrical disruption experienced by customers. For each day $d$ and ZCTA $i$, the outage intensity $I_{i,d}$ was computed as

$$I_{i,d} = \frac{1}{n_{i,d}} \sum_{t=1}^{n_{i,d}} P_{i,t,d} \tag{1}$$

where $P_{i,t,d}$ is the percentage of customers without power during time interval $t$, and $n_{i,d}$ is the number of 15-minute intervals meeting the 0.1 % threshold on that day.

**(2) Outage duration**

Outage duration quantifies the cumulative time that each ZCTA experienced power outage during a given day. This metric reflects the persistence of power service disruption over time. The outage duration $D_{i,d}$ was calculated as

$$D_{i,d} = 0.25 \times n_{i,d} \tag{2}$$

where $n_{i,d}$ is the count of 15-minute intervals in which the 0.1 % outage criterion was satisfied. Multiplying by 0.25 converts the count to hours.

**4.1.2 Food access metrics**

Daily food access metrics were derived from the Spectus and SafeGraph datasets. The food access metrics capture distinct behavioral dimensions of food access and disruption. All metrics were calculated separately for each ZCTA and day within the study period.

**(1) Redundancy**

Redundancy represents the number of distinct food POIs visited by residents from ZCTA $i$ on day $d$.

$$R_{i,d} = \text{set (food POIs visited from ZCTA } i \text{ on day } d) \tag{3}$$

Higher redundancy indicates greater availability and diversity of accessible food sources and thus higher system flexibility under disrupted conditions.

**(2) Frequency**

The metric of frequency quantifies the daily number of food-related visits normalized by the redundancy of destinations, reflecting the overall activity level of food access trips.

$$F_{i,d} = \frac{Volume_{i,d}}{R_{i,d}} \tag{4}$$

where $Volume_{i,d}$ is the total number of trips from ZCTA $i$ to all food POIs on day $d$.

**(3) Proximity**

Proximity measures the mean Euclidean distance between the centroid of each home ZCTA $i$ and all food POIs $j$ visited on a given day $d$.

$$distance_{i,j} = 2r \times sin^{-1}\left(\sqrt{sin^2\left(\frac{\varphi_{d_j} - \varphi_{c_i}}{2}\right) + cos\varphi_{d_j} cos\varphi_{c_i} sin^2\left(\frac{\lambda_{d_j} - \lambda_{c_i}}{2}\right)}\right) \tag{5}$$

where $r$ is Earth's radius (6,378.137 km); $\varphi_{c_i}$, $\lambda_{c_i}$ and $\varphi_{d_j}$, $\lambda_{d_j}$ represent the latitude and longitude of the ZCTA centroid $i$ and food POI $j$, respectively.

The daily proximity value was then calculated as

$$P_{i,d} = \frac{\sum_j distance_{i,j} \times Volume_{i,j,d}}{R_{i,d}} \qquad (6)$$

Smaller $P_{i,d}$ values indicate shorter average travel distances and therefore greater spatial accessibility.

To better characterize overall accessibility conditions and the operational continuity of food facilities, we also calculated the shortest distance and facility inactivity to support baseline comparisons and descriptive analyses. Daily comparisons of these metrics between the baseline period and the disruption period reveal both the magnitude and heterogeneity of food access disruptions across communities.

**(4) Shortest distance**

The shortest distance metrics captures the minimum Euclidean distance between the centroid of each ZCTA and its nearest food POI. This baseline metric reflects fundamental spatial accessibility to the closest available food source and is useful for comparing community-level disparities in geographic access.

**(5) Facility inactivity**

Facility inactivity assesses the operational continuity of food POIs by identifying those that experienced major declines in visits. A food facility was classified as inactive if its visit count dropped by ≥90 % relative to its mean daily baseline visits. This metric captures facilities that experienced significant functional disruption, enabling the identification of critical food facilities during and after Hurricane Beryl.

## 4.2 Analytical methods

### 4.2.1 Time-series lagged correlation

To explore the temporal coupling and delayed effects of power outages on food accessibility, we conducted a lagged correlation analysis between daily outage and food access metrics. For each ZCTA $i$, Pearson correlation coefficients $r$ were calculated between the daily time series of outage metrics and the daily time series of food access metrics:

$$r(\tau) = corr(X_{i,d}, Y_{i,d+\tau}) \tag{7}$$

where $X_{i,d}$ represents the power outage metric on day $d$, $Y_{i,d+\tau}$ is the food access metric at lag $\tau$, and $\tau$ ranges from 0 to $k$ days (e.g., 0-, 1-, 2-, 3-day lags).

Pearson correlation coefficients range from -1 to 1, where negative values indicate inverse relationships and positive values indicate direct relationships, with larger magnitudes reflecting stronger associations. The maximum absolute correlation coefficient across lags indicates the time delay at which food access responds most strongly to power outages.

### 4.2.2 Time-series clustering

To identify spatial patterns of disaster impact and recovery, time-series clustering was independently applied to the power outage metrics and food access metrics at the ZCTA level. This approach groups ZCTAs that exhibit similar temporal trajectories of disruption and restoration, enabling the recognition of typologies that characterize the heterogeneity of system responses across the Houston metropolitan area.

A three-day moving average was first applied to each ZCTA's daily time series to smooth noise and highlight underlying trends. Next, the DTW-based K-means algorithm was employed because it captures similarities in the shape of temporal patterns. DTW calculates the optimal nonlinear alignment between two time series, allowing flexible comparison even when their trajectories are phase-shifted or stretched in time (Aghabozorgi, Shirkhorshidi et al. 2015). For two time series $p = (p_1, p_2, \ldots, p_m)$ and $q = (q_1, q_2, \ldots, q_n)$, DTW identifies the minimum cumulative distance along a warping path $\pi$ through the distance matrix $D$, where each element $d_{xy} = (p_x - q_y)^2$. The algorithm minimizes the total cost of alignment while satisfying boundary, continuity, and monotonicity constraints. The K-means procedure then partitions all time series into $k$ clusters by minimizing the sum of DTW distances between each series and its assigned cluster centroid:

$$argmin_{C_1, \ldots, C_k} \sum_{i=1}^{k} \sum_{x \in C_i} \text{DTW}(x, \mu_i)^2 \tag{8}$$

where $\mu_i$ is the DTW-based centroid of cluster $C_i$. Implementation was carried out using the tslearn Python library, employing the *k-means++* initialization to enhance convergence stability and reduce sensitivity to random starting points.

The optimal number of clusters ($k$) was then determined using the silhouette score to ensure both statistical robustness and interpretive usefulness (Li and Mostafavi 2024). The score was computed to evaluate intra-cluster cohesion and inter-cluster separation, defined as

$$s(i) = \frac{b(i) - a(i)}{\max(a(i), b(i))} \tag{9}$$

where $a(i)$ is the mean DTW distance between time series $i$ and other members in its cluster, and $b(i)$ is the minimum average distance to the nearest neighboring cluster. Higher silhouette scores indicate more compact and well-separated clusters.

# 5. Results

## 5.1 Overview of power outage and food access disruption

The Houston metropolitan area contains 3,187 identified food facilities, including grocery stores, restaurants, and other retail food establishments (Figure 2a). These facilities collectively form a dense urban food access network that sustains residents' daily needs. Figure 2b illustrates typical food access patterns on July 1, representing the network of daily movements from residential ZCTAs to food POIs. The extensive interconnections across the region underscore the critical role of food access in supporting everyday mobility, household welfare, and urban system stability.

Following the landfall of Hurricane Beryl on July 8, 2024, both the power supply and food facility accessibility experienced severe disruption. Figure 2c depicts the daily number of inactive food facilities between July 6 and July 14. The number of inactive facilities increased sharply after the landfall, rising from 72 (2.26% of the total) on July 6 to 504 (15.81%) on July 8. Inactivity remained elevated on July 9 (460 facilities, 14.43%) and July 10 (357 facilities, 11.20%) before declining thereafter. This pattern indicates that a substantial portion of the region's food infrastructure was non-operational for several consecutive days, constraining residents' access to essential resources during the storm's peak and early recovery period.

The spatial distribution of cumulative outage duration from July 6 to July 20 is shown in Figure 2d. The outage impact was widespread and highly heterogeneous across the region. 35% of the 140 ZCTAs (49 ZCTAs) experienced prolonged outages exceeding 10 days, while 22 ZCTAs faced outages lasting 8~10 days, 14 ZCTAs between 6~8 days, 4 ZCTAs between 6~8 days, 41 ZCTAs between 2~4 days, and only 9 ZCTAs restored service within 2 days. These results reveal the extensive and persistent nature of the power crisis that affected large portions of the Houston area. The prolonged outages not only disrupted the operation of food facilities but also limited residents' mobility and their ability to reach functioning establishments.

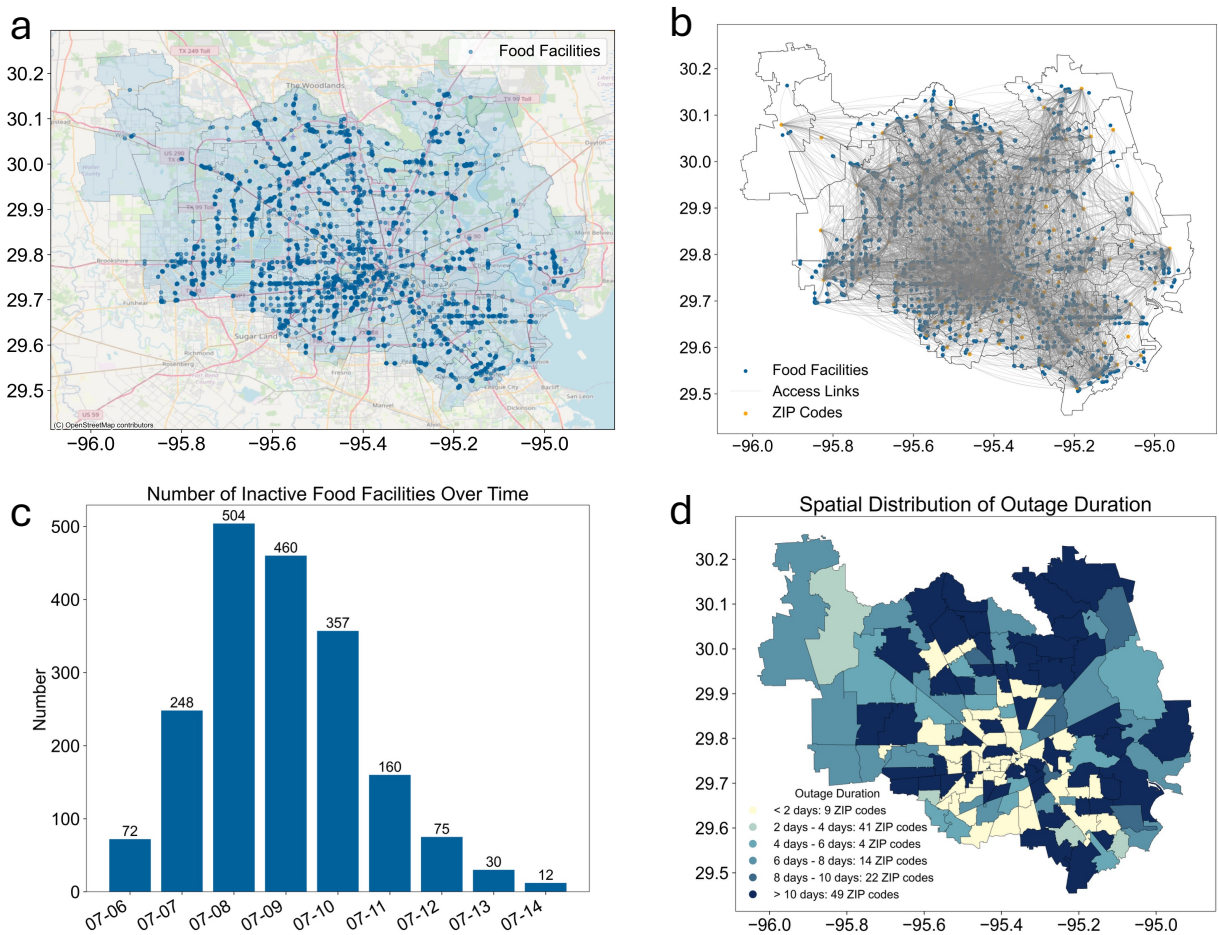

**Figure 2. Overview of food facilities and power system disruption during Hurricane Beryl.** **(a)** Spatial distribution of 3,187 food facilities in the Houston metropolitan area. **(b)** Representative baseline food access network showing movements from residential ZCTAs to food POIs on July 1. **(c)** Daily number of inactive food facilities between July 6 and July 14. **(d)** Spatial distribution of cumulative outage duration across ZCTAs during the disruption period.

**5.2 Disparities in food access during baseline and disruption periods**

To examine how different communities adapted their food access behaviors during Hurricane Beryl, we analyzed changes in the normalized shortest distance to visited food facilities between the baseline and disruption periods. The ZCTAs were classified into low and high groups based on the mean values of road density and median household income. Figure 3a presents the disparities by road density. During the baseline period, both low and high road-density ZCTAs exhibited similar median normalized shortest distances (0.034 and 0.037, respectively). After the landfall, a clear shift toward shorter trips emerged in both groups, reflecting residents' adaptation to constrained mobility and localized food access options. The decline was particularly pronounced in low road-density ZCTAs, where the median shortest distance fell from 0.034 to 0.015. This indicates a collective behavioral adjustment toward nearby food sources, likely driven by road closures, limited connectivity, or heightened safety concerns. In high road-density ZCTAs, the median shortest distance decreased from 0.037 to 0.026, showing a similar but less drastic contraction. Although both groups reduced travel, the residual distance remained higher in the high road-density areas, suggesting relatively greater accessibility flexibility within better-connected networks.

Figure 3b compares disparities across income groups. Both low and high household income communities exhibited reduced travel distances during the disruption period. For low-income areas, the median normalized shortest distance declined from 0.018 to 0.014; for high-income areas, from 0.014 to 0.009. While both groups adapted by shortening their trips, the high-income group consistently maintained shorter travel distances during both normal and disruption periods, implying higher baseline accessibility and greater spatial resilience of nearby food resources.

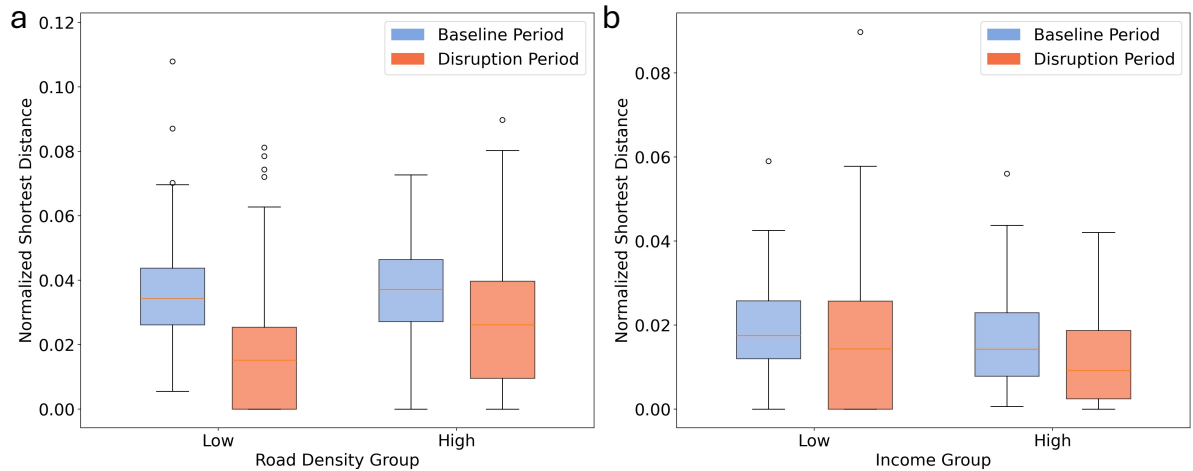

**Figure 3. Distribution of normalized shortest distance by road density and median household income groups during the baseline and disruption periods.** (**a**) Differences between low and high road-density groups. (**b**) Differences between low- and high-income groups. ZCTAs were classified into low and high groups based on mean values of road density and median household income. Statistical significance between groups was tested using one-way ANOVA ($p < 0.001$).

### 5.3 Temporal lagged impact of power outages on food access

To examine the evolving relationship between power system disruption and residents' ability to access food, we analyzed their temporal dynamics throughout the study period. Before the hurricane's approach, all three food access metrics (i.e., redundancy, frequency, and proximity) remained stable near their baseline averages (Figures 4 A1-A3). In the days preceding landfall, however, a consistent increase was observed across all three metrics. This surge reflects heightened food-seeking and stocking behavior, as residents increased visits to multiple locations in preparation for the impending hurricane. As Hurricane Beryl made landfall on July 8, these metrics exhibited abrupt and substantial declines. Redundancy and proximity dropped to nearly 0.0, indicating that the diversity of food locations visited collapsed and residents limited their travel to the nearest available food facilities. The metric of frequency also fell sharply, showing that overall food trips decreased as mobility and business operations halted. A rapid rebound occurred almost immediately after the landfall, with values rising between July 9 and July 14 before gradually stabilizing. This pattern suggests a surge of concentrated trips to the earliest reopened or most accessible food facilities once safety conditions permitted. Such pre- and post-disruption behavioral intensification has been documented in our previous work and reflects an adaptive response aimed at reinforcing household resilience (Ma and Mostafavi 2025). In contrast, both power outage intensity and duration remained minimal before July 7 but increased sharply afterward (Figures 4 B1-B2). Outage intensity and duration reached normalized values near 1.0 by July 10, approximately two days after the nadir of food access. Although a gradual decline followed, outage levels remained far above baseline conditions through July 20, indicating that electrical restoration lagged substantially behind the recovery of mobility and food-access behaviors.

The visual alignment of Figure 4A and 4B points to a temporal lag: the most severe disruption to food access occurred on July 8, whereas the peak of the outage crisis occurred around July 10. To quantify this relationship, Pearson correlation coefficients were calculated between daily power outage metrics (i.e., intensity and duration) and food access metrics (i.e., redundancy, frequency, and proximity) across lags of 0~7 days (Figure 4C). All correlations within the first three lag days were negative and statistically significant ($p < 0.01$), confirming that worsening power outages were associated with diminished food accessibility. The strength of these negative correlations consistently peaked at a lag of two days, demonstrating a delayed influence of power service disruption on food access recovery.

These results reveal a clear two-day lagged compound effect between power outages and food access recovery. While the immediate decline in food access on July 8 was driven primarily by direct physical impacts of the hurricane (e.g., flooding, wind damage, and precautionary closures) the subsequent deepening of the power outage crisis on July 10 imposed a secondary and prolonged constraint. As many food facilities required electricity to resume operations and households relied on power for mobility, refrigeration, and communication, delayed restoration of the grid directly hindered the normalization of food access behaviors.

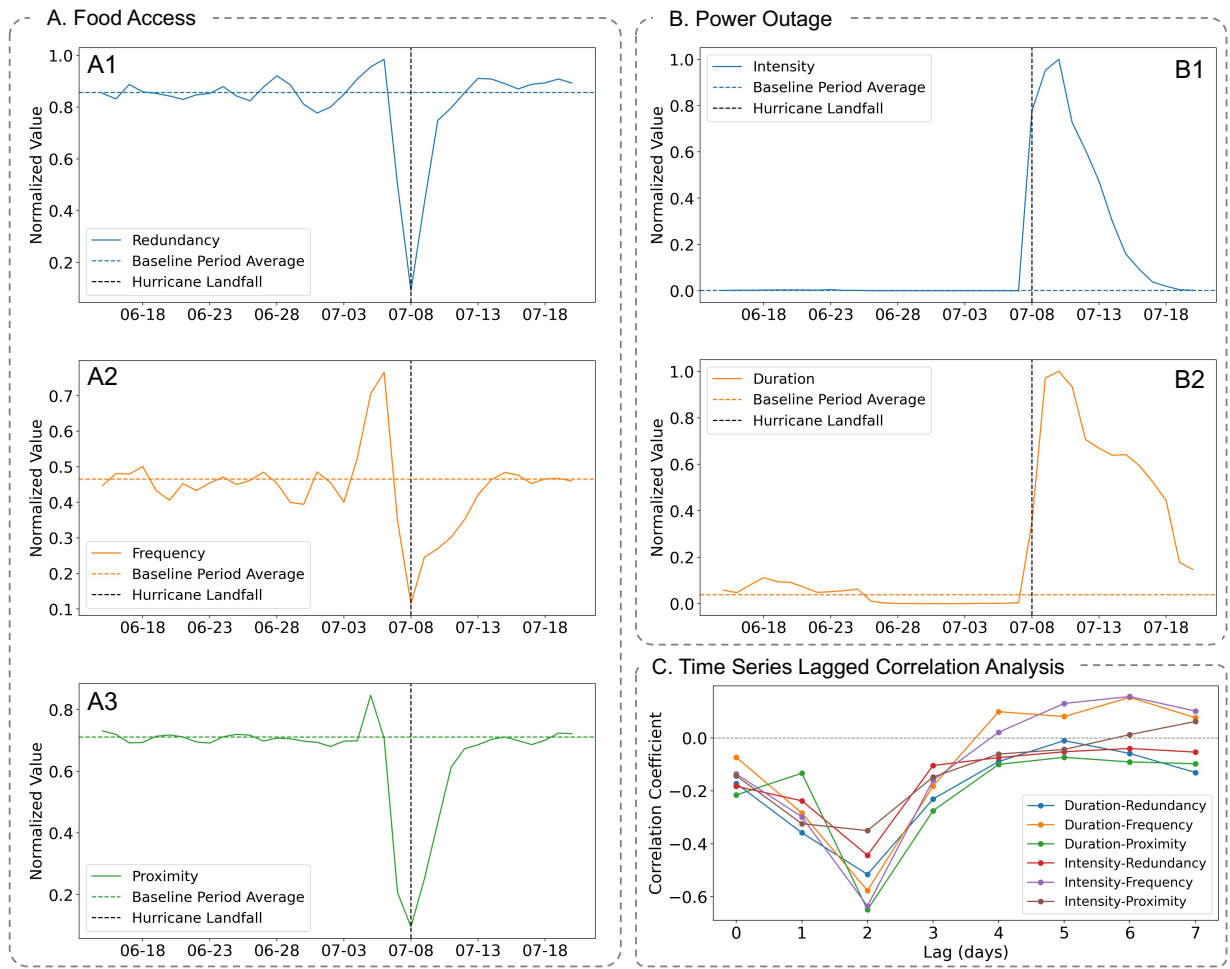

**Figure 4. Temporal coupling of power outages and food access during Hurricane Beryl. (A)** Normalized daily time series of the food access metrics: (A1) Redundancy, (A2) Frequency, and (A3) Proximity. **(B)** Normalized daily time series of the power outage metrics: (B1) Intensity, and (B2) Duration. **(C)** Time-series lagged correlations between power outage metrics and food access

metrics at 0~7-day lags. Correlations were evaluated for significance at $p < 0.01$ using two-tailed tests; correlations lost statistical significance beginning on day 4.

**5.4 Spatial distribution of compound power outages and food access disruption**

To identify distinct spatial clustering patterns of power outages and food access disruption, DTW-based K-means clustering algorithm was independently applied to the multivariate food access metrics (i.e., redundancy, frequency, and proximity) and the bivariate power outage metrics (i.e., intensity and duration) for each ZCTA during the disruption period (July 6 to July 20, 2024). Each time series was expressed as the percentage change relative to its mean value calculated over the baseline period (June 15 to July 5, 2024). The optimal number of clusters for both the food access and power outage analysis was determined to be two, as indicated by the highest silhouette scores (Table 1).

**Table 1. Model performance and optimal number of clusters ($k$) for time-series clustering of food access and power outage metrics.**

| Number of Clusters ($k$) | Food Access Metrics | Power Outage Metrics |
|---|---|---|
| 2 | 0.47 | 0.44 |
| 3 | 0.38 | 0.36 |
| 4 | 0.33 | 0.32 |
| 5 | 0.30 | 0.29 |
| 6 | 0.27 | 0.26 |

The clustering of food access time series produced two dominant patterns (Figure 5A). The cluster 0 (n = 56 ZCTAs, low food access disruption) experienced a decline in all three metrics on July 8 but demonstrated a strong and rapid rebound within days of landfall. These areas are mainly located in the western, northwestern, and some eastern peripheral zones, suggesting more adaptive recovery behavior and faster restoration of retail functionality. The cluster 1 (n = 84 ZCTAs, high food access disruption) endured a deeper and more prolonged disruption, with average redundancy values remaining well below pre-event levels for an extended period. Spatially, these ZCTAs are concentrated in central Houston, corresponding to denser urban cores where mobility restrictions and prolonged facility closures likely compounded accessibility challenges.

A similar two-cluster pattern emerged for power outage dynamics (Figure 6B). The cluster 0 (n = 24 ZCTAs, high power outage) exhibited severe electrical disruption, with outage intensity peaking several days after landfall before gradually recovering. These ZCTAs are predominantly situated in the central part of the region. The cluster 1 (n = 116 ZCTAs, low power outage) showed comparatively minor and short-lived interruptions and encompasses the majority of the Houston metropolitan area.

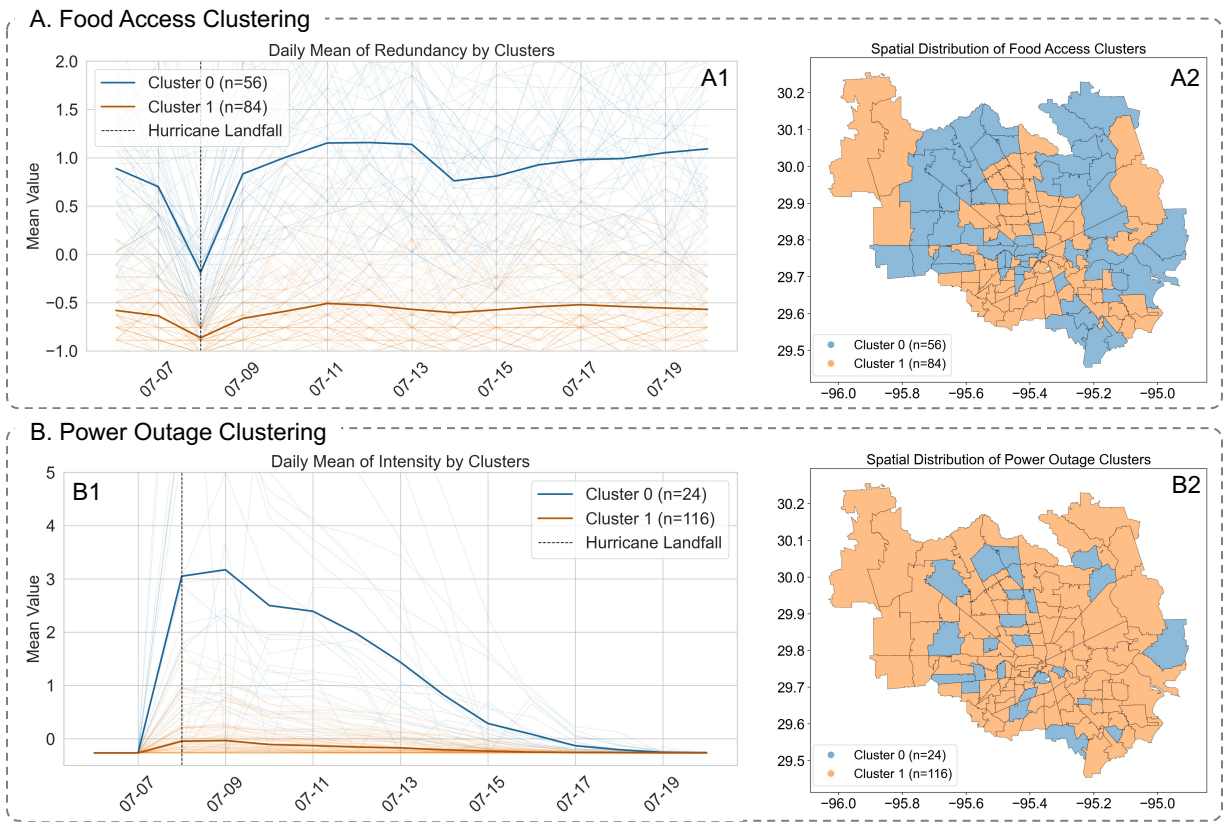

**Figure 5. Spatiotemporal distribution of food access and power outage dynamics during Hurricane Beryl. (A)** Food access clustering. (A1) Temporal trajectories of mean redundancy values by clusters. (A2) Spatial distribution of food access clusters across ZCTAs. **(B)** Power outage clustering. (B1) Temporal trajectories of mean outage intensity by clusters. (B2) Spatial distribution of power outage clusters across ZCTAs. The light-colored lines represent daily mean values for each individual ZCTA, while the bold lines depict the average trajectory across all ZCTAs within each cluster.

Overlaying the food access and power outage clustering results produced four compound typologies that jointly describe the degree of infrastructure disruption and behavioral impact across the Houston metropolitan area (Figure 6a). The first group (high power outage & high food access disruption, n = 10 ZCTAs) represents the most critically affected zones, primarily located in the northwestern and southeastern portions of the region. These areas experienced both severe power failures and high levels of food access disruption, reflecting acute infrastructural vulnerability and limited community adaptive capacity. The second group (high power outage & low food access disruption n = 14) includes ZCTAs located in the central-northwestern and central-southeastern regions. Despite significant power interruptions, these communities maintained relatively stable food access conditions or recovered quickly, indicating stronger resilience mechanisms, which may be potentially due to diversified facility networks, higher functional redundancy, or more robust socioeconomic resources. The low power outage & high food access disruption group (n = 70) encompasses neighborhoods where power disruptions were comparatively limited but food access disruption remained substantial. This pattern suggests that factors beyond power restoration (e.g., socioeconomic disadvantage, lower facility density, or pre-existing accessibility constraints) played a key role in constraining recovery. These areas highlight the importance of addressing non-infrastructure drivers of vulnerability in post-disaster planning. Finally, the low power outage & low food access disruption group (n = 46) represents the most resilient communities, predominantly located in peripheral and suburban regions. These areas experienced minimal power interruptions and relatively low food access disruption, reflecting both infrastructural robustness and higher adaptive capacity under compounding stressors.

To further characterize these typologies, distributions of road density and median household income were compared across the four groups. Figure 6b shows that clusters with greater food access disruption (i.e., the groups of high power outage & high food access disruption and low power outage & high food access disruption) exhibits lower median road density, compared with the less disrupted clusters (i.e., the groups of high power outage & low food access disruption and low power outage & low food access disruption). This finding reinforces the role of limited transportation connectivity as a structural constraint on access resilience: communities with sparser road networks are more susceptible to prolonged food access disruption during and after major power outages.

By contrast, the analysis of median household income reveals comparatively minor variations across typologies, with slightly higher incomes observed in the two high-disruption clusters (i.e., the groups of high power outage & high food access disruption and low power outage & high food access disruption) relative to the others. This finding suggests that income alone did not govern food access resilience during Hurricane Beryl. Instead, it may reflect differences in mobility behavior and consumption networks. Residents in higher-income areas often engage in broader and more spatially diverse food access patterns. Consequently, when disruptions occur, these areas experience larger relative declines in access metrics due to a greater share of their usual destinations becomes temporarily inaccessible. In contrast, lower-income communities that rely more consistently on nearby essential stores may show smaller relative disruptions, despite lower overall resource capacity.

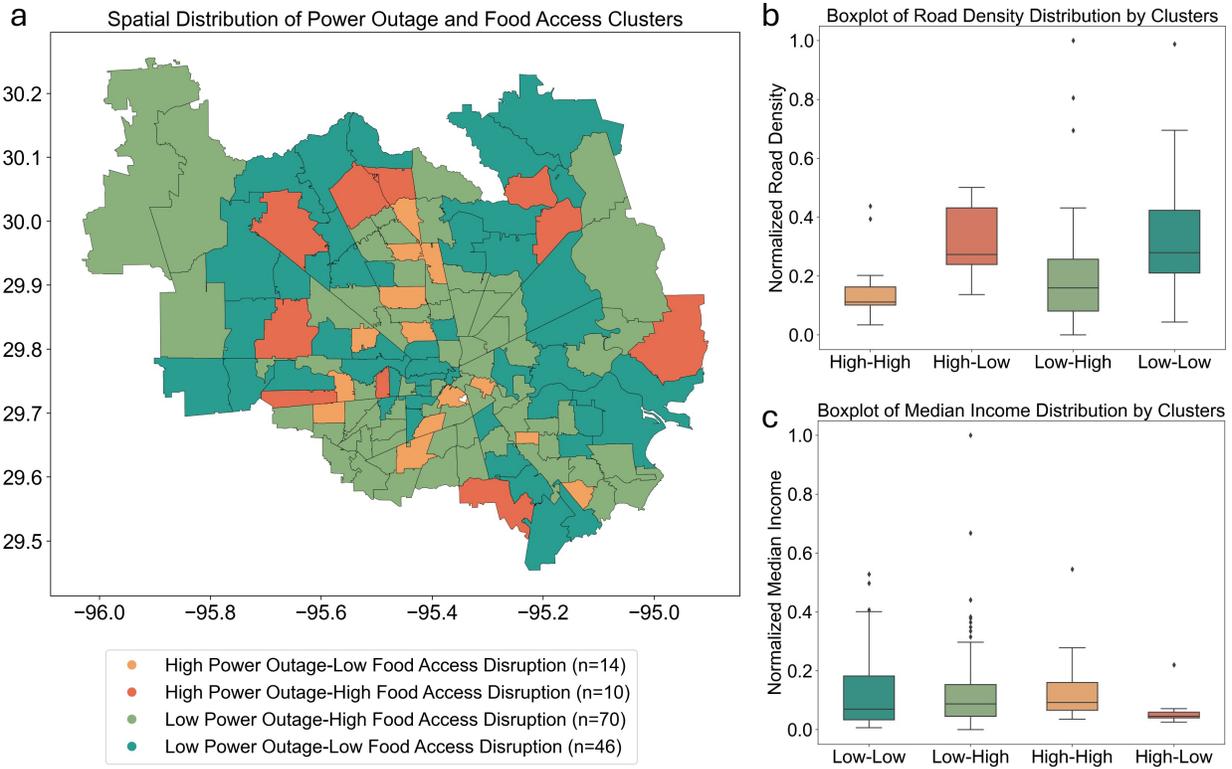

**Figure 6. Compound typologies of power outage and food access disruptions and their socioeconomic characteristics. (a)** Spatial distribution of the four compound typologies based on the overlay of power outage and food access clusters. **(b)** Boxplot of normalized road density across the four compound typologies. **(c)** Boxplot of normalized median household income across the four compound typologies. Statistical significance across groups was tested using one-way ANOVA ($p < 0.001$).

## 5.5 Identification of critical food facilities

To determine the food locations most essential for sustaining community resilience, this section identifies critical food facilities whose continued functionality is vital for maintaining community food access and equity before, during, and after disasters. Based on the preceding analyses, three criteria were applied to identify these critical facilities. The spatial distribution of the identified critical food facilities across the Houston metropolitan area is shown in Figure 7.

- **Criterion 1:** Facilities serving high power outage & high food access disruption areas

These facilities (shown as red nodes in Figure 7) are considered critical because they serve communities that experienced the compounded impact of severe power outages and pronounced food access disruption, as identified in Section 5.4. Specifically, food locations were selected if they were situated within the ten ZCTAs classified as high power outage & high food access disruption and ranked in the top 25th percentile of mean daily visits during the baseline period. A total of 92 facilities met this criteria, spatially concentrated in the central and southeastern region. These facilities represent essential access points that faced significant operational challenges but remain vital for highly impacted and energy-dependent communities.

- **Criterion 2:** Facilities serving marginalized communities

The second category (shown as dark-blue nodes in Figure 7) captures facilities located in or serving marginalized communities, defined here as the 35 ZCTAs in the lowest 25th percentile of road density distribution. Low road density constrains accessibility even under normal conditions and can severely impede mobility when transportation networks are disrupted. Within these ZCTAs, facilities ranking in the top 25th percentile of baseline mean daily visits were selected, yielding 131 locations. Spatially, these sites are primarily distributed along major transportation corridors in peripheral and semi-urban neighborhoods. Supporting these facilities is crucial to ensuring equitable food access for residents facing chronic spatial and infrastructural disadvantages.

- **Criterion 3:** Facilities experiencing major operational downtime

The third group (shown as green nodes in Figure 7) comprises food facilities that experienced significant operational disruption during the hurricane. Facilities were classified as inactive if their average daily visits decreased by 90% or more during the disruption period relative to the baseline. Among these inactive facilities, those in the top 25th percentile of pre-disaster visitation were identified as critical due to their central role in the food network prior to the event, totaling 139 locations. Spatially, these facilities are concentrated along major transportation corridors in peripheral parts of the metropolitan area, similar to those identified under Criterion 2. Understanding the factors behind their temporary failure (e.g., dependence on electricity, localized flooding, supply-chain interruptions, or workforce shortages) can inform strategies to bolster the operational resilience of similar facilities.

After accounting for overlaps across criteria, a total of 294 unique critical food facilities were identified. Collectively, these three categories offer a multi-layered perspective on food system resilience, spanning both social vulnerability and infrastructural fragility. Analyzing these cases can guide targeted mitigation investments to enhance regional resilience and reduce post-disaster inequality.

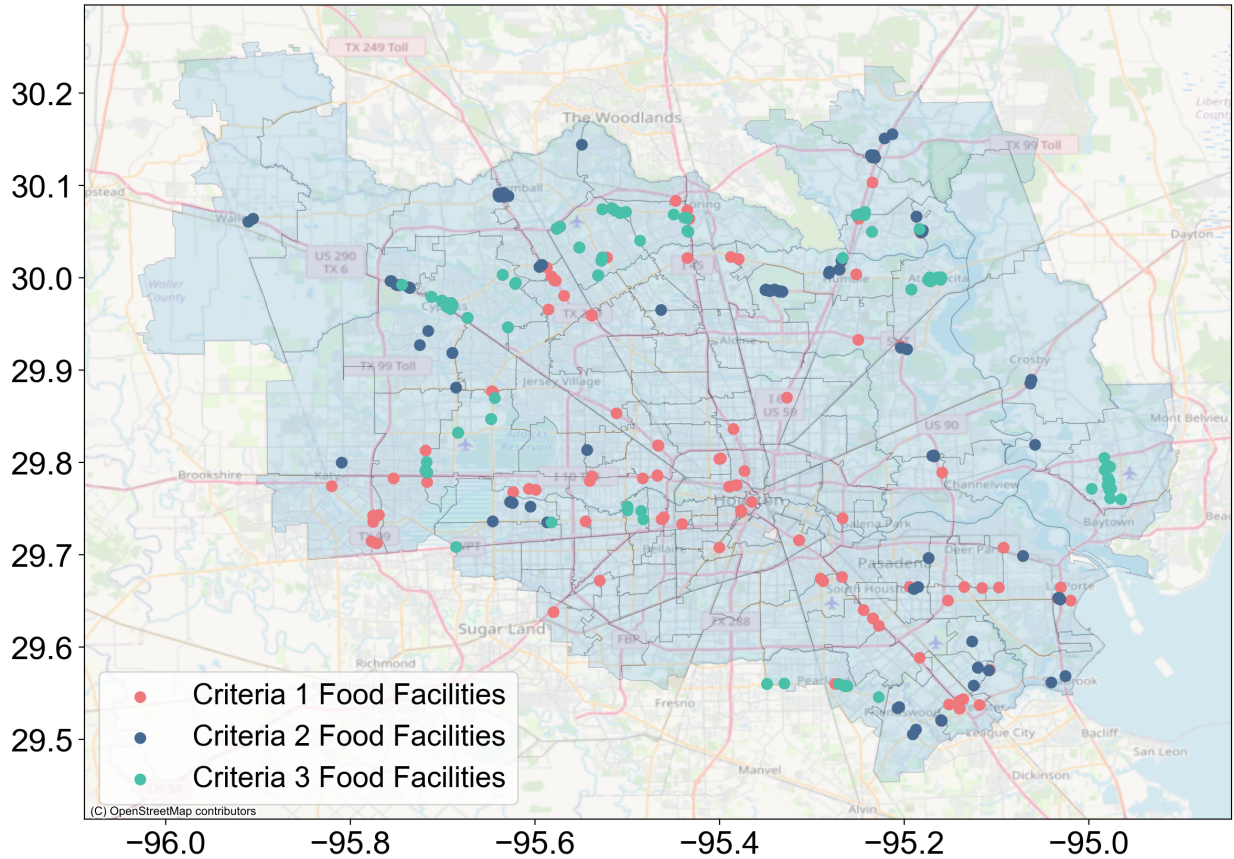

**Figure 7. Spatial distribution of the identified critical food facilities.**

## 6. Discussion and concluding remarks

This study investigated the complex interplay between power outages, human mobility, and urban food accessibility in the Houston metropolitan area following the 2024 Hurricane Beryl. By leveraging high-resolution time series outage records and POI-based mobility data, we developed daily metrics of power disruption and food access, quantified their temporal coupling and lag, uncovered spatiotemporal compound typologies, and identified critical food facilities for targeted resilience planning. Our results reveal a multi-layered picture of disaster impact and recovery that is governed by both infrastructure interdependencies and behavioral adaptation.

## 6.1 Discussion of findings

First, the analysis of food access disparities highlighted differentiated behavioral adaptations during Hurricane Beryl. Residents across all communities tended to visit closer food locations during the hurricane period. Such concurrent reduction in visitation redundancy, frequency, and proximity reflects a shift from discretionary to essential mobility, consistent with prior research on mobility-based lifestyle decline during disasters (Ma, Li et al. , Rajput, Li et al. 2022, Rajput and Mostafavi 2023, Ma and Mostafavi 2025). Further analysis based on high-low compound typologies comparisons indicates this adaptive contraction of mobility was most pronounced in areas with low road density, underscoring the role of transportation infrastructure as a structural determinant of access resilience. Road density, a key dimension of urban form and structure, has been shown in numerous studies to influence post-disaster mobility and service accessibility (Li, Yan et al. 2024, Ma, Blessing et al. 2024, Ma and Mostafavi 2024). In this study, communities with sparser road networks experienced greater and more prolonged food-access disruption, as limited connectivity restricted rerouting options and hindered recovery once major corridors were impaired. This finding reinforces the critical importance of pre-existing transportation infrastructure in shaping communities' adaptive capacity and recovery potential (Snaiki, Wu et al. 2020, Li, Wang et al. 2025). By contrast, while lower-income communities consistently exhibit higher baseline vulnerability in food access, the clustering analysis suggests that income disparities were less decisive in explaining behavioral adaptation. Higher-income communities, which typically depend on a wider range of food destinations, experienced larger relative declines in access metrics during the hurricane. Lower-income neighborhoods, more reliant on proximate essential stores, demonstrated relatively smaller proportional disruptions despite limited resource capacity. These patterns illustrate that socio-economic vulnerability and urban structural limitations influence resilience through distinct mechanisms: the former through exposure and resource constraints, and the latter through dependency breadth and spatial flexibility.

Second, our analysis identified a clear temporal lagged impact of power outages on food access. While food access metrics reached their lowest point on the day of landfall, primarily due to the direct physical impacts such as flooding, wind damage, and precautionary closures, the peak of power outage severity occurred two days later. The strongest negative correlations between the power outage metrics and the food access metrics were consistently found at this two-day lag. This suggests that while the immediate storm causes acute access disruptions, the subsequent and prolonged failure of the power infrastructure exerts a significant, delayed suppressive effect on the recovery of food access. This delay is not only statistically significant but also operationally meaningful. According to the facility inactivity metrics, 2.35% of food facilities remained non-operational as late as July 12. As communities attempt to recover, the ongoing power crisis becomes a barrier to accessing essential food resources. This temporal offset highlights a critical interdependency between infrastructure functionality and essential service accessibility. Previous disaster-recovery studies have often focused on single-system recovery trajectories, overlooking how failures in one infrastructure network delay restoration across others (Yuan, Farahmand et al. 2023, Klasa, Trump et al. 2025, Shtob, Fox et al. 2025). The presenting findings in this study empirically demonstrate that such cross-system dependencies introduce temporal inertia into the recovery process. Consequently, disaster-recovery strategies must address not only the immediate consequences of physical hazards but also the lagged cascading effects of interdependent infrastructure failures. Ensuring timely power restoration is therefore pivotal for accelerating the recovery of essential community services such as food accessibility.

Third, the development of spatial compound typologies and the identification of critical food facilities not only provided a nuanced understanding of both food access disruption and power outage severity but also offered actionable targets for intervention. By applying advanced time-series clustering algorithm to the temporal trajectories of power outage and food access metrics across ZCTAs, this study classified communities into four compound typologies that capture the heterogeneity of disaster impacts and recovery processes. The resulting combinations illustrate starkly different vulnerability conditions. The high-high areas represent areas of acute compounded vulnerability, where extensive infrastructure failure coincided with persistent access disruption, while the low-low areas denote more resilient regions that recovered rapidly after the event. These typologies advance traditional single-factor classifications by explicitly integrating infrastructure and behavioral dimensions to shape community resilience (Nickdoost, Jalloul et al. 2024, Stotten 2024). Building on these typologies, the study further identified 294 critical food facilities based on three principles: (1) service to highly impacted areas, (2) service to marginalized communities, and (3) operational fragility during the disaster. These facilities represent critical nodes within the community food network, whose functionality is essential for maintaining local food security during crises. These three categories integrate both social vulnerability and infrastructural fragility dimensions to offer a multi-layered perspective on food system resilience. This high-resolution, facility-level identification of criticality and risk moves beyond prior studies that focused primarily on classifying vulnerable areas, offering a more granular and actionable basis for disaster management, targeted investment, and resilience planning (Yuan, Farahmand et al. 2023, Li and Mostafavi 2025).

**6.2 Study contributions, significances, and implications**

Theoretically, this study first advances the theoretical understanding of compound and cascading disaster processes by revealing how the failure of one critical infrastructure system produces lagged and suppressive effects on another. Through the integration of high-frequency outage and mobility data, the study demonstrates that community resilience is not a static attribute of place or population but a dynamic condition shaped by the evolving interplay among infrastructure functionality, urban form, and human adaptation. The two-day lag between outage peaks and food access disruptions highlights the temporal coupling between physical infrastructure and behavioral response. The result also suggests that system recovery trajectories depend on the speed and synchronization of interdependent urban subsystems restoration. These findings enrich existing theories of infrastructure interdependencies, human mobility dynamics, and multi-system urban resilience. On the other hand, the results extend theoretical perspectives on human mobility and adaptive behavior under crisis conditions. By quantifying how residents modify their travel for essential needs across pre-event and post-event phases, this study empirically captures the behavioral rhythms of preparation, shock, and recovery. The observed pre-disaster surge, post-disaster overshoot, and subsequent stabilization reflect a predictable temporal logic of adaptive response rather than purely reactive behavior. This insight reframes resilience as a co-evolving product of structural connectivity and behavioral adaptation, moving beyond conventional static and physical system-based indicators of vulnerability. Collectively, these contributions provide a new lens for examining human-infrastructure interactions and advancing the theoretical discourse on multi-system urban resilience.

Methodologically, this paper establishes an operational blueprint for measuring interdependency by coupling approximately 173,000 15-minute outage observations with over 1.25 million revealed trips to 3,187 food POIs across 140 ZCTAs. The analysis derives daily indices for both outage characteristics (intensity and duration) and access metrics (redundancy, frequency, and proximity) to quantify cross-system timing rather than static co-variation. The lagged-correlation design uncovers a consistent two-day offset between the nadir of food access on July 8 and peak outage burden around July 10. The time-series clustering reveals four compound typologies when overlaid. The analysis further demonstrates that road network sparsity organizes the persistence and depth of access disruption more strongly than income once dynamics are accounted for. Together, these design choices transform the literature from parallel narratives of behavior and infrastructure into a single, event-resolved measurement system that is portable to other lifelines and hazards.

The significance of this work is twofold. First, the study introduces a time-aware empirical framework, comprising daily indices, cross-correlation over lags, and shape-based clustering, that yields testable claims about the asynchronous mechanics of impact and recovery across coupled systems. For instance, the finding that power lags access by two days creates a foundation for comparative inference across events and cities, embedding interdependency into resilience theory as timed coupling rather than coincidence. Second, by linking ZCTA-level dynamics to facility-level criticality, the work resolves scale mismatches that often blunt inference and policy translation in interdependency studies. Researchers can replicate this template to determine whether observed lags arise from infrastructure constraints, behavioral adaptation, or network topology, which can advance causal hypotheses and strengthen the external validity of lifeline-coupling research.

From a practical perspective, this paper translates cascading-effects analytics into immediate levers for response and investment. The quantified two-day dependency window identifies when delayed power restoration will continue suppressing food access even after roads reopen, which pinpoint a critical period for staging generator fleets, mobile refrigeration, and feeder-level restoration serving dense food nodes and nearby neighborhoods. The compound typologies provide a map-based prioritization surface. The identified 294 critical food facilities converts analysis into site-specific actions including backup power deployment, microgrid installation, continuity planning, and surge logistics. The work demonstrates precisely how and where outage cascades propagate into lifeline deprivation, enabling operations to intervene at the right nodes and times. The significance lies in reframing power restoration as a lifeline enabler rather than an isolated engineering objective. Sequencing decisions on the grid become determinants of household food security and the pace of social recovery. By foregrounding cascading effects from power to facility operability and human access, the study equips utilities, emergency managers, and planners with a common, data-driven target function: reducing the duration and spatial footprint of access droughts. Concretely, agencies can prioritize feeders intersecting high power outage & high food access disruption areas, pre-position backup generation for facilities in low road-density ZCTAs where alternatives are scarce, and track recovery using the same daily indices to trigger all-clear signals or surge support. These interventions accelerate recovery while promoting equity, and the framework readily transfers to other lifelines such as water and communications where outage cascades similarly condition essential service access.

### 6.3 Limitations and future directions

There were also limitations in this study, which could be addressed in the future. First, both the mobility and outage datasets are subject to representational and spatial constraints. The Spectus mobility data rely on smartphone users who opt in to share location information, potentially underrepresenting certain demographic groups such as elderly residents. Similarly, the ZCTA-level power outage data, while the most detailed available, may obscure finer neighborhood-level variations. Future research should incorporate higher-resolution datasets (e.g., feeder-level outage telemetry, community surveys) to better capture localized access dynamics and demographic diversity. Second, although the study identifies strong lagged correlations between power outages and food access disruptions, these results do not establish causality. Unobserved factors such as localized flooding, road closures, or communication failures may also shape both power restoration and behavioral adaptation. Future research should employ multi-hazard and multi-infrastructure analyses and integrate additional hazard layers and network data to clarify causal mechanisms and assess compound effects more holistically.

## Data availability

The human mobility data that used in this study are available from Spectus, Inc. and SafeGraph, Inc., but restrictions apply to the availability of these data, which were used under license for the current study. The power outage data can be obtained from ORNL upon request. Other data we used in this study are publicly available.

## Code availability

All analyses were conducted using Python. The code that supports the findings of this study is available from the corresponding author upon request.

## Acknowledgements

This work was supported by the National Science Foundation under Grant CMMI-1846069 (CAREER). Any opinions, findings, conclusions, or recommendations expressed in this research are those of the authors and do not necessarily reflect the view of the funding agency.

## Competing interests

The authors declare no competing interests.